\documentstyle[12pt]{article}

\font\sevenbf=cmbx7
\def\note#1]{{\sevenbf #1 ------]}}
\begin{document}

\begin{center}

{\bf On density and temperature-dependent
ground-state and continuum effects in the
equation of state for stellar interiors --\\
A Comment on the paper by S. Arndt, W. D\"appen and
A. Nayfonov 1998, ApJ 498, 349}

\vspace{0.5cm}

Wolf-Dietrich Kraeft$^1$, Stefan Arndt$^2$,
Werner D\"appen$^{3,4}$, Alan Nayfonov$^{3,5}$

\vspace{0.2cm}

$^1$ Institut f\"ur Physik, Universit\"at Greifswald,
D-17489 Greifswald, Germany \\
$^2$ Max-Planck-Institut f\"ur Plasmaphysik,
D-17489 Greifswald, Germany\\
$^3$ Department of Physics and Astronomy,
University of Southern California, Los Angeles, CA 90089-1342, U.S.A\\
$^4$ Theoretical Astrophysics Center, Institute for
Physics and Astronomy, Aarhus Uni\-versity, 8000 Aarhus C, Denmark\\
$^5$ IGPP, Lawrence Livermore National Laboratory, Livermore, CA 94550,
U.S.A.\\

\end{center}

\newpage

\noindent
{\bf Abstract}

\noindent
Misunderstandings have occurred regarding the conclusions of the paper
by S. Arndt, W. D\"appen and A. Nayfonov 1998, ApJ 498, 349. At
occasions, its results were interpreted as if it had shown basic flaws
in the general theory of dynamical screening. The aim of this comment
is to emphasize in which connection the conclusions of the paper have to
be
understood in order to avoid misinterpretations.

\vspace{0.3cm}

\newpage

\begin{center}

{\bf Comment}

\end{center}

The paper ``On density and temperature-dependent ground-state and
continuum effects in the equation of state for stellar interiors'' by
Arndt, D\"appen and Nayfonov (1998; hereinafter ADN) dealt with the
consequence of density and temperature effects on two-particle
properties (such as binding energies and energies of continuum) for the
equation of state, under conditions relevant for stellar interiors.\\
The ADN paper applied, among other, the work by Seidel, Arndt and Kraeft
(1995; hereinafter SAK). In particular, ground-state and continuum-edge
shifts under plasma conditions are contained in SAK. In the ADN paper,
an empirical approach is used to obtain thermodynamic quantities from
the data contained in the SAK paper, in view of their possible testing
by helioseismology.\\
Essentially, two results of the SAK calculations were used by ADN. One
was based on an elaborate formalism for dynamic screening, the other on
a static approximation. The result of the ADN paper was that when the
raw data of the SAK figure were taken to be thermodynamic quantities,
plausible results (labeled STATIC) emerged for the static continuum
edge, but the same technique yielded rather absurd quantities (labeled
DYNAMIC) for the corresponding dynamic continuum edge.\\
Although the conclusions of ADN were worded to allow various
possibilities for the interpretation of this outcome, especially
pointing at a potential inadequacy of the empirical identification of
two-particle properties and thermodynamic quantities, it has been
brought to our attention that the ADN paper is still at occasions
misunderstood as if it had shown basic flaws in the general theory of
dynamical screening.\\
The aim of this comment is to explain once again in which connection the
conclusions of ADN have to be understood in order to avoid
misinterpretations, and to stimulate further investigations.\\

\vspace{0.3cm}

The determination of two-particle properties is outlined in paragraph
2.3 of ADN along the lines given in the references
(Zimmermann et al. 1978; Kraeft et
al. 1986) and SAK, and takes carefully into account especially the
dynamic character of the effective interaction and of the self energy.
In paragraph 2.4.2 of ADN, the influence of bound and continuum states
on thermodynamic functions is considered. Bound state energies are only
very weakly density dependent (at least for $Z=1$ ions), consequently
we
will discuss here only the continuum edge problem. The continuum edge is
defined as the sum of (momentum dependent) single particle energies, or
of the self energies, respectively, taken for zero momenta.
Consequently, these continuum energies are two-particle quantities, not
thermodynamic ones.\\
ADN were considering various approximation levels, referred to as DEBYE,
STATIC and DYNAMIC. The results on the first two approximation levels
differ only very little from each other as the Hartree-Fock
contributions in Eq. (21) of ADN are not important for the densities
considered. The Debye-shift [last term of Eq. (21)] is the result of
thermodynamic averaging of the momentum dependent self energies over a
Maxwellian and thus a thermodynamic quantity. This was shown explicitly
in Kraeft et al. (1986), p. 115. Therefore, the Debye-shift
gives the
interaction part of the chemical potential (and thus the internal energy
contribution $U_4$) in the Debye case. \\
At the level of the dynamical approximation of the self energies, as
considered in SAK and taken as a part of the approximation level DYNAMIC
in ADN, the analogous thermodynamic averaging was not performed.
ADN have chosen a simple procedure to estimate a
thermodynamic quantity; this approximation consists in taking the
continuum lines according to SAK. It leads to excellent thermodynamic
quantities in the DEBYE and STATIC cases, but to severe discrepancies
with observationally admissible results in the DYNAMIC case.\\
In this comment, we want to emphasize that the unphysical result of the
DYNAMIC approximation is due to the absence of thermodynamic averaging
in the continuum curves on the static and dynamic approximation levels,
respectively.
By being an already thermodynamically averaged result, the Debye curve
gives a physically consistent result. (By mere coincidence, the
static case gives reasonable results, too.)
However, such a consistency is
absent when the dynamic continua according to SAK are used as
thermodynamic data. \\
In general, of course, a rigorous theory always has to account for the
dynamics before doing (thermodynamic) averages. Of course, at the
thermodynamic level, after averaging, the dynamics is no longer
perceptible. \\
While we have stressed here the importance of an inconsistency in the
approximation introduced by ADN, their paper remains so far the only
attempt to perform explicitly a quantitative calculation of
thermodynamic quantities in the presence of dynamical screening. Future
studies will have to show how to include such effects more rigorously.\\

\vspace{0.3cm}

{\it Acknowledgments:} W.-D.~K. acknowledges support by
the Deutsche For\-schungsgemeinschaft, Sonderforschungsbereich 198.
W.~D. and A.~N. acknowledge support by the grant
AST-9618549 of the National Science Foundation and the SOHO Guest
Investigator grant NAG5-6216 of NASA. W.~D. was supported in part by the
Danish National Research Foundation through its establishment of the
Theoretical Astrophysics Center.
\\

\end{document}